\documentstyle [12pt]{article}
\begin{document}
\title{Non-Exponential Decay for Polaron Model}
\author{L. Accardi, S.V. Kozyrev, I.V. Volovich}

\maketitle

\centerline{\it Centro Vito Volterra Universita di Roma Tor Vergata}

\begin{abstract}{
A model of particle interacting with quantum field is considered.
The model includes as particular cases the polaron model
and non-relativistic quantum electrodynamics.
We compute matrix elements of the evolution operator
in the stochastic approximation and show that
depending on the state of the particle one can get the
non-exponential decay with the rate $t^{-{3\over 2}}$.
In the process of computation a new algebra of commutational relations
that can be considered as an operator deformation of quantum Boltzmann
commutation relations is used.
}
\end{abstract}

\section{Introduction}

For many dissipative systems one has the exponential time decay
of correlations. This result was established for various models
and by using various approximations, see for ex. \cite{Loui}.
For certain models, in particular for the spin-boson Hamiltonian,
also a regime with the oscillating behavior was found
\cite{Leggett1}, \cite{Leggett2}, \cite{AcKoVol}.
The presence of such a regime is very important in the ivestigation
of quantum decoherence.
The aim of this note is to show that for the model of particle interacting
with quantum field, in particular for the polaron model,
one can have not only the standard exponential decay
but also the non-exponential decay (as some powers of time) of correlations.

We investigate the model describing interaction of non-relativistic particle
with  quantum field.
This model is widely studied in elementary particle physics,
solid state physics, quantum optics, see for example
\cite{Bog}-\cite{Feynman}.
We consider the simplest case in which matter is represented by a
single particle, say an electron, whith position and momentum
$q=(q_{1}, q_{2}, q_{3})$ and
$p=(p_{1}, p_{2}, p_{3})$  satisfying the commutation relations
$ [q_{j}, p_{n} ] = i \delta_{jn}$. The electromagnetic field is described by
boson operators
$a(k)=(a_{1}(k), a_{2}(k), a_{3}(k));a^{\dag}(k)=
(a^{\dag}_{1}(k), \ldots ,a^{\dag}_3(k))$
satisfying the  canonical commutation relations
$[a_{j}(k),a_{n}^{\dag}(k')]=\delta_{jn}\delta(k-k')$.
The Hamiltonian of a free non relativistic atom interacting with a quantum
electromagnetic field is
\begin{equation}\label{1.1}
H=H_0+\lambda H_I=\int\omega(k)a^{\dag}(k)a(k)dk+{\frac{1}{2}}\,p^2+\lambda
H_I
\end{equation}
where $\lambda$ is a small constant, $\omega (k)$ is the dispersion law
of the field,
\begin{equation}\label{1.2}
H_I=\int d^3k\left(g(k)
p\cdot a^{\dag}(k)  e^{-ikq} +
\overline g(k)p\cdot a(k)e^{ikq}  \right) + h.c.
\end{equation}
Here $p\cdot a(k)=\sum_{j=1}^{3}p_j a_j(k)$, $p^2=\sum_{j=1}^{3}p_j^2$,
$a^{\dag}(k)a(k)=\sum_{j=1}^{3}a^{\dag}_j(k)a_j(k)$, $kq=\sum_{j=1}^{3}k_j
q_j$.

For the polaron model the Hamiltonian has the form
$$
H=\int\omega(k)a^{\dag}(k)a(k)dk+{\frac{1}{2}}\,p^2+\lambda
\int d^3k\left(g(k)
a^{\dag}(k)  e^{-ikq} +
\overline g(k) a(k)e^{ikq}  \right)
$$
It is different from (\ref{1.1}), (\ref{1.2}) by a momentum $p$
in the interaction Hamiltonian. For the analysis of this paper
this difference is not important.

In the present paper we will use the  method
for the approximation of the quantum mechanical evolution
that is called the stochastic limit method, see for example
\cite{AcKoVol}, \cite{VanHove}-\cite{AcLuVo}.
The general idea of the stochastic limit is to make the time rescaling
$t\to t/\lambda^2$ in the solution of the Schr\"o\-din\-ger equation in
interaction picture $U^{( \lambda )}_t=e^{itH_0} e^{-itH}$,
associated to the Hamiltonian $H$, i.e.
$$
{ \frac{\partial}{ \partial t} }  U^{( \lambda )}_t
=-i { \lambda  } H_I (t) \ U_t^{(\lambda )}
$$
with $H_I(t)=e^{it H_0}H_Ie^{-itH_0}$.
We get the rescaled equation
$$
{ \frac{\partial}{  \partial t} }  U^{( \lambda )}_{t/\lambda^2}
=- {\frac{i}{\lambda}} H_I (t/\lambda^2) \
U_{t/\lambda^2}^{(\lambda )}
$$
and one wants to study the limits, in a topology to be specified,
\begin{equation}\label{1.6}
\lim_{\lambda\to0}U^{(\lambda)}_{t/\lambda^2}= U_t;\qquad
\lim_{\lambda\to0}{\frac{1}{\lambda}}\,H_I\left({\frac{t}{\lambda^2}}\right)
=H_t
\end{equation}
We will prove that $U_t$ is the solution of the equation
\begin{equation}\label{1.7}
\partial_t U_t\,=-iH_t U_t\quad;\qquad U_0=1
\end{equation}
The interest of this limit equation is in the fact that many problems
become explicitly integrable.
The stochastic limit of the model (\ref{1.1})-(\ref{1.2})
has been considered in \cite{AcLu}, \cite{AcLuVo}, \cite{Gou96},
\cite{Ske96}, \cite{q-alg}, \cite{tmf}.

After the rescaling $t\to t/\lambda^2$ we consider the
simultaneous limit $\lambda\to 0$, $t\to\infty$ under the condition
that $\lambda^2 t$ tends to a constant (interpreted as a new {\it slow scale}
time). This limit captures the main
contributions to the dynamics in a regime, of {\it long times and small
coupling\/} arising  from the cumulative effects, on a large time scale,
of small interactions ($\lambda\to 0$). The physical idea is that,
looked from the slow time scale of the atom, the field looks like a very
chaotic object: a {\it quantum white noise}, i.e. a $\delta$-correlated
(in time) quantum field $b_j^{\dag}(t,k), b_j(t,k)$ also called a {\it master
field}.  If one introduces the dipole approximation the master field is the
usual boson Fock white noise. Without the dipole approximation the master
field is described by a new type of commutation relations
of the following form
\cite{AcLuVo}
\begin{equation}\label{1.8}
b_j(t,k)p_n=(p_n-k_n)b_j(t,k)
\end{equation}
\begin{equation}\label{1.9}
b_j(t,k)b_n^{\dag}(t',k')=2\pi\delta(t-t')
\delta\left(\omega(k)-kp+\frac{1}{2}k^2\right)
\delta(k-k')\delta_{jn}
\end{equation}
Such quantum white noises can be treated as an operator deformation of quantum
Boltzmann commutation relations.
Recalling that $p$ is the particle momentum, we see that
the relation (\ref{1.8})
shows that the particle and the master field are not independent even at a
kinematical level. This is what we call {\it entanglement}.
The relation (\ref{1.9}) is a generalization of the algebra of free
creation--annihilation operators with commutation relations
$$A_iA^{\dag}_j=\delta_{ij}$$
and the corresponding statistics becomes a generalization of the
Boltzmannian (or Free) statistics. This generalization is due to
the fact that the right hand side is
not a scalar but an operator (a function of the atomic momentum). This
means that the relations (\ref{1.8}), (\ref{1.9}) are {\it module commutation relations}.
For any fixed value $\bar p$ of the atomic momentum we get a copy of the
free (or Boltzmannian) algebra. Given the relations
(\ref{1.8}), (\ref{1.9}), the
statistics of the master field is uniquely determined by the condition
$$
b_j(t,k) \Psi = 0
$$
where $\Psi $ is the vacuum of the master field, via a module
generalization of the free Wick theorem, see \cite{q-alg}.

In  Section 2 the dynamically $q$-deformed commutation relations
(\ref{2.5}), (\ref{2.6}), (\ref{2.4}) are
obtained  and the stochastic limit for collective operators is evaluated.
In Section 3 the stochastic limit of the evolution equation  is found.
In Section 4 the non-exponential decay for vacuum vector
in the polaron model is investigated.

\section{Deformed commutation relations}

In this section we reproduce in the brief form the notations and
the main results of the work \cite{q-alg}.

In order to determine the limit (\ref{1.6}) one rewrites
the rescaled interaction Hamiltonian in terms of some rescaled fields
$a_{\lambda,j}(t,k)$:
$$
{\frac{1}{\lambda}}\,H_I\left({\frac{t}{\lambda^2}}\right)=
\int d^3k p(\overline g(k) a_\lambda(t,k)+  g(k)a^{\dag}_\lambda(t,k)) + h.c.
$$
where
$$
a_{\lambda,j}(t,k):={\frac{1}{\lambda}}\,e^{i{\frac{t}{\lambda^2}}\,H_0}
e^{ikq}a_j(k)e^{-i{\frac{t}{\lambda^2}}\,H_0}=
\frac{1}{\lambda}e^{-i{\frac{t}{\lambda^2}}\,
\left(\omega(k)-kp+\frac{1}{2}k^2\right)}e^{ikq}a_j(k)
$$
It is now easy to prove
that operators $a_{\lambda,j}(t,k)$ satisfy the following $q$--deformed
module relations,
$$
a_{\lambda,j}(t,k)a^{\dag}_{\lambda,n}(t',k')=
$$
\begin{equation}\label{2.5}
=a^{\dag}_{\lambda,n}(t',k')a_{\lambda,j}(t,k)
\cdot q_\lambda(t-t',kk')+
{\frac{1}{\lambda^2}}\,q_\lambda\left(t-t',\omega(k)-kp+\frac{1}{2}k^2\right)
\delta(k-k')\delta_{jn}
\end{equation}
\begin{equation}\label{2.6}
a_{\lambda,j}(t,k)p_n=(p_n-k_n)a_{\lambda,j}(t,k)
\end{equation}
where
\begin{equation}\label{2.3}
q_\lambda(t-t',x)=e^{-i{\frac{t-t'}{\lambda^2}}\,x}
\end{equation}
is an oscillating exponent.
This shows that the module $q$--deformation of the commutation relations arise
here as a result of the dynamics and are not put artificially {\it ab
initio}.
For a discussion of $q$-deformed commutation relations see for example
\cite{AreVo}.
Now let us suppose that the master field
\begin{equation}\label{2.7}
b_j(t,k)=\lim_{\lambda\to0}a_{\lambda,j}(t,k)
\end{equation}
exist. Then it is natural to conjecture that its algebra shall be
obtained  as the stochastic limit ($\lambda\to0$) of the
algebra  (\ref{2.5}), (\ref{2.6}).
Notice that
the factor $q_\lambda(t-t',x)$ is an oscillating exponent and one easily sees
that
\begin{equation}\label{2.8}
\lim_{\lambda\to0}q_\lambda(t,x)=0\ ,
\qquad\lim_{\lambda\to0}{\frac{1}{\lambda^2}}\,
q_\lambda(t,x)=2\pi\delta(t)\delta(x)
\end{equation}
Thus it is natural
to expect that the limit of (\ref{2.6})
is
\begin{equation}\label{2.9}
b_j(t,k)p_n=(p_n-k_n)b_j(t,k)
\end{equation}
and the limit of (\ref{2.5}) gives the module free relation
\begin{equation}\label{2.10}
b_j(t,k)b_n^{\dag}(t',k')=2\pi\delta(t-t')
\delta\left(\omega(k)-kp+\frac{1}{2}k^2\right)
\delta(k-k')\delta_{jn}
\end{equation}
Operators $a_{\lambda,j}(t,k)$ also obey the relation
\begin{equation}\label{2.4}
a_{\lambda,j}(t,k)a_{\lambda,n}(t',k')=a_{\lambda,n}(t',k')a_{\lambda,j}(t,k)
q_\lambda^{-1}(t-t',kk')
\end{equation}
In what follows we will not write indexes $j$, $n$ explicitly.
The
relation (\ref{2.4}) should disappear after the limit, see \cite{q-alg}.
In fact, if the relation
(\ref{2.4}) would survive in the limit then,
because of (\ref{2.8}), it should give
$b(t,k)b(t',k')=0$, hence also $b^{\dag}(t,k)b^{\dag}(t',k')=0$, so all the
$n$--particle vectors with $n\geq2$ would be zero.

\section{ Evolution equation}

Let us find stochastic differential equation for the model we consider.
In the introduction we claimed that the stochastic limit for the
Shr\"odinger equation in interaction picture will have the form (\ref{1.7}):
$\partial_t U_t\,=-iH_t U_t$.
But in this equation  both $H_t$ and $U_t$ are distributions.
We need to regularize this product of distributions.
In the present section we will make the following  regularization:
roughly speaking we replace $H_t$ by $H_{t+0}+const$.

We investigate the evolution operator in interaction picture
$U^{(\lambda)}_t$.
We start with the equation
$$U^{(\lambda)}_{t+dt}=\biggl(1+(-i\lambda)\int^{t+dt}_tH_I(t_1)
dt_1+
$$
$$
+(-i\lambda)^2\int^{t+dt}_tdt_1\int^{t_1}_tdt_2H_I(t_1)
H_I(t_2)+\dots\biggr)U^{(\lambda)}_t
$$
where $dt>0$.
We get for $dU^{(\lambda)}_t=U^{(\lambda)}_{t+dt}-U^{(\lambda)}_t$
$$
dU^{(\lambda)}_t=\biggl((-i\lambda)\int^{t+dt}_tH_I(t_1)dt_1+
(-i\lambda)^2\int^{t+dt}_tdt_1\int^{t_1}_tdt_2H_I(t_1)H_I(t_2)+
\dots\biggr)U^{(\lambda)}_t
$$
Let us make the rescaling $t\to t/\lambda^2$ in this perturbation theory
series. We get
$$dU^{(\lambda)}_{t/\lambda^2}=\biggl((-i)\int^{t+dt}_tdt_1{1\over
\lambda}\,H_I\left({t_1\over\lambda^2}\right)+
$$
\begin{equation}\label{rescalseries}
+(-i)^2\int^{t+dt}_t
dt_1\int^{t_1}_tdt_2\,{1\over\lambda}\,H_I\left({t_1\over\lambda^2}
\right)\,{1\over\lambda}\,
H_I\left({t_2\over\lambda^2}\right)+\dots\biggr)U^{(\lambda
)}_{t/\lambda^2}
\end{equation}
To find the stochastic differential equation we need
to collect all the terms of
order $dt$ in the perturbation theory
series (\ref{rescalseries}).
Terms of order  $dt$  are contained only in the first two terms of these
series.  Let us investigate the first two terms.
For the first term of the perturbation theory we get
\begin{equation}\label{first_term}
\int^{t+dt}_tdt_1{1\over\lambda}\,H_I\left({t_1\over\lambda^2}
\right)=
\int^{t+dt}_tdt_1
\int dk\left(\overline g(k)(2p+k)a_\lambda(t_1,k)+g(k)a^{\dag}_\lambda
(t_1,k)(2p+k)\right)
\end{equation}
In the stochastic limit $\lambda\to0$ this term gives us
$$\int dk\left(\overline g(k)(2p+k)dB(t,k)+g(k)dB^{\dag}(t,k)(2p+k)\right)
$$
where the stochastic differential $dB(t,k)$
is the stochastic limit of the field $a_\lambda(t,k)$
in the time  interval $(t,t+dt)$:
$$
dB(t,k)=\lim_{\lambda\to 0}
\int_{t}^{t+dt}d\tau\,  a_\lambda(\tau,k)=\int_{t}^{t+dt}d\tau\, b(\tau,k)
$$
We will prove that the stochastic differential
$dB(t,k)$  and the evolution operator
$U_t$ are  free
independent.  In the bosonic case independence would result in the relation
$[dB(t,k), U_t]=0$.
From this relation follows that
$\langle X\, dB(t,k) U_t\rangle =0$ for arbitrary observable $X$.
In the case of Boltzmannian statistics  we get the same relation:
the (free) independence
means  that roughly speaking $dB(t,k)$ kills all creations in $U_t$.
We have the following

{\it Lemma.\qquad}{\sl
The stochastic differental  $dB(t,k)$
and the evolution operator $U_t$ are  free independent.
This means that for  an arbitrary observable $X$
$$
\langle X\, dB(t,k) U_t\rangle =0\quad \forall X
$$
Here $\langle\cdot\rangle$ is the stochastic limit of the
vacuum expectation of boson field
(that acts as a conditional expectation on momentum of quantum particle $p$).
}

We will prove this result by analizing of the perturbation theory series.
We have
$$
\langle X\, dB(t,k) U_t\rangle=\lim_{\lambda\to 0}
\langle X_{\lambda} \int_{t}^{t+dt}d\tau\,  a_\lambda(\tau,k)
\biggl(1+(-i)\int^{t}_0 dt_1\, {1\over\lambda}\,
H_I\left({t_1\over\lambda^2}\right)
+ \dots+
$$
$$ +(-i)^n\int^{t}_0 dt_1\dots \int^{t_{N-1}}_0 dt_N\,
{1\over\lambda}\,H_I\left({t_1\over\lambda^2}\right)\dots
{1\over\lambda}\,H_I\left({t_n\over\lambda^2}\right)+\dots
\biggr)\rangle
$$
where ${1\over\lambda}\,H_I\left({t_k\over\lambda^2}\right)$
is given by the formula (\ref{first_term}).
Here $\lim_{\lambda\to 0}X_{\lambda}=X$.
Let us analize the $N$-th term of perturbation theory.
The  $N$-th term of perturbation theory
is the linear combination of the following terms
(we omit integration over $k$, $k_n$)
$$
\langle X_{\lambda} \int_{t}^{t+dt}d\tau\,  a_\lambda(\tau,k)
\int^{t}_0 dt_1\dots \int^{t_{N-1}}_0 dt_N\,
a^{\varepsilon_1}_\lambda(t_1,k_1)\dots
a^{\varepsilon_N}_\lambda(t_N,k_N)\rangle
$$
Let us shift $a_\lambda(\tau,k)$ to the right using dynamically $q$-deformed
relations. In the following we will use notions of the work \cite{q-alg}.
Let us enumerate annihilators in the product
$
a^{\varepsilon_1}_\lambda(t_1,k_1)\dots
a^{\varepsilon_N}_\lambda(t_N,k_N)
$
as $a_\lambda(t_{m_j},k_{m_j})$, $j=1,\dots J$, and enumerate
creators as  $a^{\dag}_\lambda(t_{m'_j},k_{m'_j})$, $j=1,\dots I$, $I+J=N$.
This means that if $\varepsilon_m=0$ then
$a^{\varepsilon_m}_\lambda(t_m,k_m)=a_\lambda(t_{m_j},k_{m_j})$
for $m=m_j$ (and the analogous condition for $\varepsilon_m=1$).

We will use the following recurrent relation for correlator
(analogous formula was proved in the work \cite{q-alg}):
$$
\langle X_{\lambda} \int_{t}^{t+dt}d\tau\,
\int^{t}_0 dt_1\dots \int^{t_{N-1}}_0 dt_N\,
a_\lambda(\tau,k)
a^{\varepsilon_1}_\lambda(t_1,k_1)\dots
a^{\varepsilon_N}_\lambda(t_N,k_N)\rangle=
$$
$$
=
\sum_{j=1}^{I}\delta ( k- k_{m'_j})
\langle X_{\lambda} \int_{t}^{t+dt}d\tau\,
\int^{t}_0 dt_1\dots \int^{t_{N-1}}_0 dt_N\,
a^{\varepsilon_1}_\lambda(t_1,k_1)\dots
{\widehat a_\lambda^{\dag}(t_{m'_j},k_{m'_j})} \dots
a^{\varepsilon_N}_\lambda(t_N,k_N)\rangle
$$
$$
\frac{1}{\lambda^2}
q_{\lambda}\left( \tau - t_{m'_j},\omega (k) -k p +\frac{1}{2}k^2\right)
\prod_{m_i>m'_j}
q_{\lambda}^{-1}\left( \tau - t_{m'_j},k k_{m_i}\right)
\prod_{m'_i>m'_j}
q_{\lambda}\left( \tau - t_{m'_j},k k_{m'_i}\right)
$$
\begin{equation}\label{recur}
\prod_{m_i<m'_j}q_{\lambda}^{-1}(\tau - t_{m_i},k k_{m_i})
\prod_{m'_i<m'_j}q_{\lambda}(\tau - t_{m'_i},k k_{m'_i})
\end{equation}
Here the notion ${\widehat a_\lambda^{\dag}}$ means that we omit the operator
$a_\lambda^{\dag}$ in this product.

The right hand side of the equation (\ref{recur}) is equal to
$$
\sum_{j=1}^{I}\delta ( k- k_{m'_j})
\langle X_{\lambda}
\int^{t}_0 dt_1\dots \int^{t_{N-1}}_0 dt_N\,
a^{\varepsilon_1}_\lambda(t_1,k_1)\dots
{\widehat a_\lambda^{\dag}(t_{m'_j},k_{m'_j})} \dots
a^{\varepsilon_N}_\lambda(t_N,k_N)\rangle
$$
$$
\frac{1}{\lambda^2}
q_{\lambda}\left(  - t_{m'_j},\omega (k) -k p +\frac{1}{2}k^2\right)
\prod_{m_i>m'_j}
q_{\lambda}^{-1}\left(  - t_{m'_j},k k_{m_i}\right)
\prod_{m'_i>m'_j}
q_{\lambda}\left(  - t_{m'_j},k k_{m'_i}\right)
$$
$$
\prod_{m_i<m'_j}q_{\lambda}^{-1}( - t_{m_i},k k_{m_i})
\prod_{m'_i<m'_j}q_{\lambda}( - t_{m'_i},k k_{m'_i})
$$
$$
\int_{t}^{t+dt}d\tau\,
q_{\lambda}\left( \tau,\omega (k) -k p +\frac{1}{2}k^2\right)
\prod_{m_i>m'_j}
q_{\lambda}^{-1}\left( \tau ,k k_{m_i}\right)
\prod_{m'_i>m'_j}
q_{\lambda}\left( \tau ,k k_{m'_i}\right)
$$
$$
\prod_{m_i<m'_j}q_{\lambda}^{-1}(\tau ,k k_{m_i})
\prod_{m'_i<m'_j}q_{\lambda}(\tau ,k k_{m'_i})
$$
(we use that $q_{\lambda}$ is an exponent). The first three lines of this
formula do not depend on $\tau$ and the last two lines do not depend
on $t_1,\dots,t_N$. Therefore the stochastic limits for these values
can be made independently (the limit of product is equal to the
product of limits). It is easy to see that the stochastic limit for
the multiplier that depends on $\tau$ (of the last two lines)
is equal to zero.   This finishes the proof of the Lemma.

The second term of the perturbation theory series is equal (up to terms
of order $(dt)^2$)
$$\int^{t+dt}_tdt_1\int^{t_1}_tdt_2{1\over\lambda}\,H_I
\left({t_1\over\lambda^2}\right){1\over\lambda}\,H_I\left({t_2
\over\lambda^2}\right)=$$
$$=\int^{t+dt}_tdt_1\int^{t_1}_tdt_2\int dk|g(k)|^2(2p+k)^2
{1\over\lambda^2}\,e^{-i{t_1-t_2\over\lambda^2}\,\left(\omega(k)-
kp+{1\over2}\,k^2\right)}$$
due to $q$--module relations on $a_\lambda(t,k)$, $p$.
Performing integration over $t_1$, $t_2$ and using the formulas
$$
\int_{t}^{t+dt}dt_1\,\int_{t}^{t_1}dt_2\,
{1\over\lambda^2}\, e^{-i\frac{t_1-t_2}{\lambda^2}x}=
\int_{t}^{t+dt}dt_1\, \int_{-t_1/\lambda^2}^{0}d\tau\,e^{i\tau x}
$$
$$
\int^0_{-\infty}dte^{itx}=\frac{-i}{x-i0}
=\pi\delta(x)-i\ P.P.{1\over x}$$
we get for the second term
$$
-i dt\int dk|g(k)|^2(2p+k)^2
{1\over\omega(k)-kp+{1\over2}\,k^2-i0}
$$
Let us denote
$$(g|g)_-(p)=
-i \int dk|g(k)|^2(2p+k)^2
{1\over\omega(k)-kp+{1\over2}\,k^2-i0}=
$$
$$
=\int dk|g(k)|^2(2p+k)^2\left(\pi\delta\left(\omega(k)-kp+{1
\over2}\,k^2\right)-i\ P.P.{1\over\omega(k)-kp+{1\over2}\,k^2}\right)
$$
Combining all the terms of order  $dt$
we get the following result:

{\it Theorem.}{\sl\qquad
The stochastic differential
equation for $U_t=\lim\limits_{\lambda\to0}U^{(\lambda)}_{t/\lambda^2}$
have a form
\begin{equation}\label{stocheq}
{dU_t}\,=\biggl(-i\int dk\left(\overline g(k)(2p+k)dB(t,k)+g
(k)dB^{\dag}(t,k)(2p+k)\right)-dt\,(g|g)_-(p) \biggr)U_t
\end{equation}
}

The equation (\ref{stocheq}) can be rewritten in the language of
distributions as
\begin{equation}\label{distrib}
\frac{dU_t}{dt}\,=\biggl(-i\int dk\left(\overline g(k)(2p+k)b(t,k)+g
(k)b^{\dag}(t,k)(2p+k)\right)-\,(g|g)_-(p) \biggr)U_t
\end{equation}
Here we uderstand the singular product of distributions $b(t,k)U_t$
in the sense that (\ref{distrib}) is equivalent to (\ref{stocheq}).
We have to stress that
$dB(t,k)=\int_{t}^{t+dt}d\tau\, b(\tau,k)\ne b(t,k)dt$
and we can not obtain (\ref{distrib}) dividing (\ref{stocheq})
by $dt$.

\section{Non-exponential decay}

Let us investigate the behavior of $\langle U_t\rangle$ using
stochastic differential equation (\ref{stocheq}). We get
$$
\langle dU_t\rangle=\langle\biggl((-i)\int dk\overline
g(k)(2p+k)dB(t,k)-dt\,(g|g)_-(p) \biggr)U_t\rangle
$$
Using the free independence of $dB(t,k)$ and $U_t$ we get
$$
{d\over dt}\,\langle U_t\rangle=\langle{d\over dt}\,U_t\rangle=-
(g|g)_-(p)\langle U_t\rangle
$$
Because $U_0=1$, we have the solution
$$\langle U_t\rangle=e^{-t(g|g)_-(p)}$$

In this section we calculate the matrix element $\langle X|U_t|X\rangle$
where $X=f(p)\otimes\Phi$ in the momentum representation and $\Phi$ is
the vacuum vector for the master field. This matrix element is equal to
\begin{equation}\label{smeared}
\langle X|U_t|X\rangle=\int dp|f(p)|^2e^{-t(g|g)_-(p)}
\end{equation}
We investigate the polaron model when $\omega(k)=1$.
For this choice of $\omega(k)$ we get
$$
\omega(k)-kp+\frac{1}{2}k^2=1-\frac{1}{2} p^2 +\frac{1}{2}(k-p)^2
$$
One can expect non-exponential relaxation when
$\hbox{ supp }f(p)\subset \{|p|<\sqrt{2}\}$.
In this case  $\hbox{Re}(g|g)_-(p)=0$ and there is no dumping.
All decay in this case is due to interferention.

We will use the approximation
$\hbox{diam supp }g(k)\gg\hbox{ diam supp }f(p)$.
Physically this means that the particle is more localized in momentum
representation than the field. This assumption seems natural because the
field's degrees of freedom are fast and the particles one are slow.
Under this assumption we can estimate the matrix element (\ref{smeared}).
We will prove that in this case there will be polynomial decay.

For $|p|<\sqrt{2}$ we get
$$
(g|g)_{-}(p)=-i\int dk\, |g(k)|^2 (2p+k)^2
\frac{1}{1-\frac{1}{2}p^2 +\frac{1}{2}(k-p)^2}=
$$
$$
=-2i \int dk\, |g(k)|^2 -i \left(I_1+I_2\right);
$$
$$
I_1=\left(-2+10p^2\right) \int dk\, |g(k)|^2
\frac{1}{1-\frac{1}{2}p^2 +\frac{1}{2}(k-p)^2}
$$
$$
I_2=6 \int dk\, |g(k)|^2   p(k-p)
\frac{1}{1-\frac{1}{2}p^2 +\frac{1}{2}(k-p)^2}
$$
Here only $I_1$ and $I_2$ depend on $p$ and therefore can interfere.
Let us find the asymptotics of $(g|g)_{-}(p)$ on $p$
(we investigate the case when $p$ is a small parameter).

We will use the following assumption on $g(k)$: let $g(k)$
be a very smooth function. This means that
$|g(k)|^2=\lambda F(\lambda k)$, $F(k)>0$
is compactly supported smooth function, $\lambda$ is a small parameter.
Let us consider the Taylor expansion on the small parameter $p$
$$
\lambda F(\lambda k)=\lambda F(\lambda (k-p)) +
\lambda^2\sum_i p_i \frac{\partial}{\partial k_i} F(\lambda (k-p)) +\dots.
$$
We get that $\lambda F(\lambda (k-p))$ is a leading term with respect to
$\lambda$. Taking $\lambda\to 0$ we get that we can use
$|g(k-p)|^2$  instead of $|g(k)|^2$  in the formulas for $I_1$ and $I_2$
for sufficiently smooth $g(k)$.
Let us calculate $I_1$ a d $I_2$. Using assumptions considered above we get
$$
I_1=\left(-2+10p^2\right) \int dk\, |g(k-p)|^2
\frac{1}{1-\frac{1}{2}p^2 +\frac{1}{2}(k-p)^2}=
$$
$$
=\left(-2+10p^2\right)
\int dk\, |g(k)|^2
\frac{1}{1+\frac{1}{2}k^2}
-p^2
\int dk\, |g(k)|^2
\frac{1}{\left(1+\frac{1}{2}k^2\right)^2}
$$
$$
I_2=pQ,\qquad Q= 6 \int dk\, |g(k)|^2   k
\frac{1}{1+\frac{1}{2}k^2}
$$
We get
$$
(g|g)_{-}(p)=
-2i \int dk\, |g(k)|^2 +
2i \int dk\, |g(k)|^2\frac{1}{1+\frac{1}{2}k^2}
-iA p^2 -i pQ;
$$
\begin{equation}\label{A}
A=10\int dk\, |g(k)|^2
\frac{1}{1+\frac{1}{2}k^2}-
\int dk\, |g(k)|^2
\frac{1}{\left(1+\frac{1}{2}k^2\right)^2}
\end{equation}
We get for $X(t)=\langle X|U_t|X\rangle$
$$
X(t)=\int dp|f(p)|^2e^{-t(g|g)_-(p)}=
$$
$$
= e^{it2\left(\int dk\, |g(k)|^2 -
\int dk\, |g(k)|^2\frac{1}{1+\frac{1}{2}k^2} \right)}
\int dp|f(p)|^2  e^{it\left(Ap^2+pQ\right)}
$$
Let us estimate this integral for $f(p)=e^{-Bp^2}$, $B>>1$.
Let us consider  for simplicity the case $Q=0$
(for example $g(k)$ is spherically symmetric).
In this case the integral is equal to
$$4\pi\int^\infty_0dp \, p^2 e^{-Bp^2}e^{iAtp^2}
=\left(\frac{\pi}{B- i At}\right)^{\frac{3}{2}}
$$
We get that for large $t$ the decay of the matrix element
$X(t)=\langle X|U_t|X\rangle$ is proportional to
$\left(At\right)^{-\frac{3}{2}}$ where $A$ is the functional of the cut-off
function given by  (\ref{A}).

To conclude, in this paper
we obtain that in the polaron model  for some (symmetric and very smooth)
cut-off functions we have the
polynomial relaxation, the matrix element being proportional to
$t^{-\frac{3}{2}}$. The dependence on the parameter $B$
that corresponds to the size of the support of the
smearing function $f(p)$ of quantum particle in the momentum space
for large  $t$ is not important. We can say that particles with large
momentum decay exponentially and the particles with small momentum
decay as $t^{-\frac{3}{2}}$ and the decay for large $t$ does not depend
on the smearing function $f(p)$.

{\bf Acknowlegements}
\medskip

S.Kozyrev and I.Volovich are grateful to Centro Vito Volterra
where this work was started for kind hospitality.
This work was partially supported by INTAS 96-0698
and RFFI-9600312 grants.
I.V. Volovich is supported in part by a fellowship
of the Italian Ministry of Foreign Affairs organized by the Landau
Network-Centro Volta.

\end{document}